%% file: bare_conf.tex
\documentclass[conference]{IEEEtran} 

\usepackage{graphicx}
\usepackage[toc]{multitoc}
\usepackage{setspace}
\usepackage{tikz}
\usepackage{pgfplots}
\usepackage{amssymb}
\usepackage{txfonts}
\usepackage[font=footnotesize]{subcaption} 
\usepackage[caption=false,font=footnotesize]{subfig} 
\usepackage[font=footnotesize]{caption} 
\usepackage{url}

\usepackage{verbatim}
\usepackage{color} 

\begin{document}
%
\title{\huge Breaking Through the Full-Duplex Wi-Fi Capacity Gain}

\author{\IEEEauthorblockN{Saulo Queiroz\IEEEauthorrefmark{1}\IEEEauthorrefmark{2}}
\IEEEauthorblockA{\IEEEauthorrefmark{1}Academic Department of Informatics \\
Federal University of Technology (UTFPR)\\ Ponta Grossa--PR, Brazil.\\
Email: sauloqueiroz@utfpr.edu.br}
\and
\IEEEauthorblockN{Jo\~ao Vilela}
\IEEEauthorblockA{Department of Informatics Engineering \\
 University of Coimbra \\ Coimbra, Portugal \\
Email: jpvilela@dei.uc.pt}
\and
\IEEEauthorblockN{Roberto Hexsel}
\IEEEauthorblockA{\IEEEauthorrefmark{2}Department of Informatics \\
 Federal University of Paran\'a (UFPR) \\ Curitiba--PR, Brazil \\
Email: roberto@inf.ufpr.br}
}

\maketitle

\begin{abstract}
In this work we identify a seminal design guideline that prevents current 
Full-Duplex (FD) MAC protocols to scale the FD capacity gain (i.e. $2$$\times$ the
half-duplex throughput) in single-cell Wi-Fi networks.
Under such guideline (referred to as $1$:$1$), a MAC protocol attempts
to initiate up to two simultaneous transmissions in the FD bandwidth. 
Since in single-cell Wi-Fi networks MAC performance is bounded by the PHY layer capacity,
this implies gains strictly less than $2\times$ over half-duplex at the MAC layer.
To face this limitation, we argue for the $1$:$N$ design guideline.
Under $1$:$N$, FD MAC protocols `see' the FD bandwidth through $N$$>$$1$ orthogonal
narrow-channel PHY layers. Based on theoretical results and software defined radio experiments, 
we show the $1$:$N$ design can leverage the Wi-Fi capacity gain more than $2$$\times$ \emph{at} 
and \emph{below} the MAC layer. This translates the denser modulation scheme incurred
by channel narrowing and the increase in the spatial reuse offer enabled by channel 
orthogonality. With these results, we believe our design guideline can inspire a new
generation of Wi-Fi MAC protocols that fully embody and scale the FD capacity gain.
\end{abstract}

\section{Introduction}
Recent works have demonstrated the feasibility of Self-Interference Cancellation 
(SIC)  techniques, turning Full-Duplex (FD) radios into a reality 
e.g.~\cite{bharadia-fullduplex-sigcomm-2013}. 
Such radios are capable of receiving and transmitting simultaneously within the
same frequency band, achieving a gain of $2\times$ the half-duplex link capacity in 
theory (i.e. the FD gain).  An important question raised by that achievement is 
whether it is possible to design a Medium Access Control (MAC) protocol that
accomplishes the goal of \emph{scaling} the FD gain in a wireless network.
A possible way to accomplish that consists in relying on the wide area implied by 
multi-cell deployments to activate multiple concurrent links~\cite{fumac-icnp-2014}. 
However, by surveying the MAC literature e.g.~\cite{goyal-distrmac-asilomar-2013, kim-janusmac-2013, 
duarte-fdlegacymac-2012,singh-contraflow-wiopt-2011}, one can find out it is hard 
to accomplish that scalability goal within a \emph{single-cell} Wi-Fi compliant Wireless Local 
Area Network (WLAN), since the contention overheads and the lack of spatial reuse 
can shrink the FD gain to $1.58$$\times$ as the network grows~\cite{xie-doesfddouble-infocom-2014}.

To tackle the limitation of current FD MAC protocols, we go a step further
and identify a common design strategy we refer to as the $1$:$1$ MAC design guideline.
With the $1$:$1$ design, an FD MAC protocol `sees' the whole FD bandwidth through a
\emph{single} PHYsical layer. To maximize FD gains with such design, MAC protocols attempt
to minimize the difference between the start time of two concurrent transmissions in the 
channel. This leads to gains bounded by the capacity of two nodes freely
transmitting to each other in the channel.
In fact, in a  single-cell WLAN, the MAC throughput is bounded by 
the PHY layer capacity.
Thus, doubling such capacity with FD radios may limit the maximum capacity gain 
achieved at the FD MAC layer to a value strictly less than $2\times$ the half-duplex
throughput. 
This suggests one needs to improve the capacity below the MAC layer more than $2\times$ to
give room for MAC protocols that actually approaches the FD gain. 


In this paper we report novel results that break through the capacity gain
leveraged by FD radios in single-cell WLANs. We accomplish this by arguing for
an alternative FD MAC design guideline we refer to  as $1$:$N$. Under that, the MAC
layer arranges the FD bandwidth  into $N$$>$$1$ PHY layers. Each PHY layer is assigned to
a portion of spectrum that is narrower than the available FD bandwidth and orthogonal
to the other PHY's spectrum portions. Similar design have been studied before from
the perspective of MAC and/or radio architecture e.g.~\cite{queiroz-sac-2015, queiroz-letters-13,
fica-ton-2013, hong-picasso-sigcomm-2012, chintalapudi-wifinc-usenix-2012}. These 
works highlight the advantages of parallel narrow channels on a single radio but 
under the half-duplex constraint. To fully realize the FD gain over a wireless bandwidth
allocated to concurrent narrow channels, one has to refer to the same kind of 
wide-band SIC design (e.g.~\cite{bharadia-fullduplex-sigcomm-2013}) assumed by current
state-of-the-art $1$:$1$ FD MAC proposals. 
We refer to such advance to report \emph{unprecedented contributions towards the FD gain 
scalability in WLANs}.

Our first contribution is to show that, contrary to the popular assumptions and beliefs,
it is possible to attain more-than-doubled capacity gains within an FD bandwidth
i.e. \emph{below} the MAC layer.
Indeed, narrowing a channel relaxes receive sensitivity requirements enabling denser
modulation schemes~\cite[Table 18--14]{ieee80211-12}.
Thus, spectrum usage improves. For instance, instead of occupying a $10$ MHz channel
with two (FD) transmissions, one can split it into two $5$ MHz orthogonal FD channels
and activate four concurrent transmissions. This yields gains of $\approx$$2.2$$\times$
over a $10$ MHz half-duplex link even considering guard-bands. We demonstrate this
theoretically and through a proof-of-concept study with USRP platforms. 

Our second contribution is to scale the novel FD gain \emph{at} the MAC layer. We characterize
the ideal condition for an $1$:$1$ FD Wi-Fi MAC protocol  and show its performance 
improves more than twice under the $1$:$N$ guideline. This happens because channel
orthogonality multiplies FD opportunities by increasing the spatial reuse offer.
We believe these results instigate further research towards a solid FD IEEE 802.11 stack.

 The remainder of this paper is organized as follows. In section~\ref{sec:systemmodel} we 
present our system model and background. In sections~\ref{sec:proposal}
and~\ref{sec:model} we present the $1$:$N$ design guideline and its capacity model, 
respectively. In section~\ref{sec:results} we present our results.   
In section~\ref{sec:conclusion} we present conclusion and future work.

\section{System Model and Background}\label{sec:systemmodel}
We consider the design directives that a Wi-Fi compliant FD MAC protocol should
follow to scale the FD gain. In this sense we focus on models to assess
capacity upper-bounds \emph{at} and \emph{below} the MAC layer in a 
single-cell infrastructure IEEE 802.11 WLAN. For the MAC
protocol study, the cell is composed of one Access Point (AP) and $n$ STAtions (STA).
STAs perform the standard CSMA/CA to initiate a transmission to the AP (uplink). The 
AP is assumed to always have a frame enqueued to its current transmitting STA. Then,
the AP can establish an FD (down)link to the STA upon processing its incoming header. 
As we discuss in section~\ref{sec:model}, this corresponds to an ideal condition
the capacity upper-bound of an FD Wi-Fi MAC protocol can be derived from.

For each MAC proposal we assume saturated traffic and ideal channel conditions~\cite{bianchi-jsac-2006}.
These assumptions ensure we assess `the most each MAC protocol can do' when
provided with best conditions. Note, however, any MAC protocol under
the design guideline we are about to present might actually perform better in noisy environments.
This happens because the narrow Wi-Fi channels we rely on are less prone to noise,
as we discuss in the section \ref{subsec:betterSNR}. Also, we assume each compared MAC 
and PHY model suffers from the same level of negligible self-interference residue. Again, 
a successful (de)modulation process might be less demanding in terms of SIC requirements 
if performed over narrower channels instead of wide channels~\cite{bharadia-fullduplex-sigcomm-2013}.

\subsection{FD MAC WLAN Terminology}
The ultimate goal of any FD MAC protocol is to take advantage of FD opportunities
within a given wireless channel to maximize capacity.
It means the protocol attempts to activate two overlapping transmissions to 
maximize channel utilization so throughput. In Wi-Fi compliant WLANs, the \emph{Primary 
Transmitter} (PT) is the first node to start transmitting a data frame after winning 
a typical CSMA/CA contention round.
The node PT transmits to is called \emph{Primary Receiver} (PR). During the primary 
transmission, the FD MAC protocol may start a secondary transmission in the channel.
In this case the sender and receiver are called \emph{Secondary Transmitter} (ST) and 
\emph{Secondary Receiver} (SR), respectively.

Basically, the FD opportunities can be classified into either \emph{symmetric} or
\emph{asymmetric}  dual-links~\cite{singh-contraflow-wiopt-2011}. In symmetric dual-links,
PT and PR coincide with SR and ST, respectively (i.e. [PT$=$SR]$\rightleftarrows$[PR$=$ST],
where the direction of each arrow denotes the destination of a transmission).
In asymmetric dual-links, there must be a \emph{third} node involved in the secondary
communication. 
Such node is either a SR or a ST. In the former case, the PR coincides with the ST i.e. 
PT$\to$[PR$=$ST]$\to$\emph{SR}. Otherwise the PT coincides with SR, i.e. 
PR$\leftarrow$[PT=SR]$\leftarrow$\emph{ST}.
Note the two possible asymmetric dual-links are not different views of the same 
scenario since in one case an already \emph{receiving} node starts transmitting while
in the other an already \emph{transmitting} node starts receiving.

\subsection{Medium Access Control Challenges with Dual-links}
The performance of an FD MAC protocol results from a balance between how
effectively it exploits dual-links and the cost it takes towards
that. Concerning asymmetric dual-links, the main challenge consists in 
assuring the secondary transmission does not collide with some possible
ongoing primary transmission. Collisions may happen whenever the
receiver node of a primary (secondary) transmission is within the 
interference range of a secondary (primary) transmission. 
In case of symmetric dual-links, the challenge consists in identifying a 
pair of nodes that have frames to each other. To maximize FD gains regardless of the 
type of dual-link, \emph{any FD Wi-Fi MAC protocol attempts to minimize}
$\Delta_t$$=$$t_{st}$$-$$t_{pt}$$\geq$$0$. Particularly for our scenario, $t_{pt}$ is the
time at which a STA starts a primary transmission after winning a CSMA/CA contention round
and $t_{st}$ is the time at which the AP starts the corresponding secondary transmission.

\subsection{Novel Classification for FD MAC Protocols}\label{sec:proposal}
In this work we identify a new category for the design of FD MAC protocols.
With this novel category, MAC protocols are classified according to the way
they exploit the available wireless FD bandwidth. In this sense, we identify
a seminal trend we refer to as the $1$:$1$ MAC design 
guideline~\cite{goyal-distrmac-asilomar-2013, kim-janusmac-2013, 
duarte-fdlegacymac-2012,singh-contraflow-wiopt-2011}. Under the $1$:$1$
guideline the MAC protocol `sees' the FD bandwidth through a \emph{single} PHY layer. 
Thus, the best-case of any $1$:$1$ MAC protocol is bounded to the capacity of a 
dual-link.
Moreover the resulting capacity is impaired because of the contention overheads.

\section{The $1$:$N$ MAC Design Guideline}\label{sec:proposal}
A reasonable way to overcome the performance limitation of $1$:$1$ FD MAC 
protocols consists in, firstly, improving the capacity below the MAC layer.
Toward that goal we advocate an alternative FD MAC design guideline we refer
to as $1$:$N$. Under this novel guideline, a MAC protocol sees the FD bandwidth 
through $N$$>$$1$ PHY layers. Each PHY layer is assigned to a sub-channel that is
narrower than the whole available FD bandwidth and orthogonal to the sub-channel
of the other PHY layers. The value of $N$ is a trade-off figure of merit between
the maximization of the number of concurrent transmissions and the minimization 
of the spectrum overhead needed to isolate channels through guard-bands. While a
comprehensive understanding about the effects of varying $N$ makes a strong 
case for future research, along this work we propose a case study for $N$$=$$2$
to quantify the unique benefits our proposal brings for the  design of FD MAC protocols. 
\subsection{Increased Spatial Reuse Offer}
The $1$:$N$ design creates more FD opportunities than $1$:$1$
by increasing spatial reuse offer, as shown in Fig.~\ref{fig:spatial-reuse-comparison}.
In the $1$:$1$ best-case scenario (Fig.~\ref{fig:1:1}) a dual-link can increase throughput
while avoiding that a hidden node (e.g. STA $S_2$) collides with an ongoing transmission 
(e.g. $S_1$$\to$$A$). However, this sacrifices spatial reuse by interfering with all other 
STAs (dashed waved arrows)~\cite{xie-doesfddouble-infocom-2014}. By arranging the FD bandwidth
into $N$ \emph{orthogonal} narrower-channel PHY layers, the $1$:$N$ best-case scenario overlaps
$N$$-$$1$ \emph{additional} dual-links in the same space. This is illustrated on Fig.~\ref{fig:1:N}
for $N$$=$$2$, in which channel orthogonality (i.e. gray and black colors) \emph{also} helps 
against collisions and enables one additional dual-link in the network.
\input{figures/spatial-reuse}
\subsection{Improved Signal to Noise Ratio} \label{subsec:betterSNR}
Prior works~\cite{arslan-acorn-ieeeton-2012, chandra-ccr-08} show experimentally that halving
a single Wi-Fi channel increases the total energy in the bandwidth, 
yielding an SNR gain of $\approx$$3$ dB. We enhance these tests to check whether the SNR statement
holds when the total active bandwidth remains the same but the number (then the width) of channels 
changes.
In each test, we set Wi-Fi signals to the same parameters. However,
one scenario considers a $10$ MHz-wide channel and the other considers two concurrent $5$ 
MHz-wide channels. 
To achieve such concurrency one can resample a $5$ MHz Wi-Fi signal by interpolating it to $2$
in the baseband. The resulting signal is duplicated and each copy is shifted to its specific 
half within a $10$ MHz band.
In Fig.~\ref{fig:psd} we plot the Power Spectrum (PS) of the strongest signals 
as reported by a couple of single-antenna Ettus USRP B210 platform. 
We estimate the PS samples and their average based on the Matlab's \texttt{pwelch} procedure.
From the plots, one can see each narrow channel benefits from  $\approx$$3$ dB gain over the
wider channel. In fact,  although both narrow channels occupy the same 
$10$ MHz spectrum, they are employed independently. Thus, both the environmental and
noise floors experienced within a channel does not account for the signal processing
in the other. 
\input{figures/psd}

\subsection{Capacity Model Below the MAC Layer}
The SNR improvements resulting from channel narrowing can translate into higher 
capacity for a Wi-Fi bandwidth. Consider an AWGN Wi-Fi channel measuring $B$ (Hz) under
a given $SNR$ (unitless). According to the Hartley-Shannon theorem,  the
maximum information that can be modulated and carried over a half-duplex bandwidth
$B$ is $C_{hd}$ Bits/Hz/s (Eq. \ref{eq:shannon}).
Assuming an FD radio and expressing the $SNR$ in dB ($SNR_{dB}=10\log(SNR)$), one
derives Eq. \ref{eq:shannondB} for the capacity limit $C_{fd1}$ of FD MAC protocols under
the $1$:$1$ design.
\begin{eqnarray}
 C_{hd} &=& B\log_2(1 + SNR)\label{eq:shannon} \\
 C_{fd1} &=& 2B\log_2\big(1 + 10^{SNR_{dB}/10}\big)\label{eq:shannondB}
\end{eqnarray}
With the $1$:$N$ guideline, the FD bandwidth $B$ is equally divided among $N$ 
narrow channels. Considering $N$$=$$2$, the $3$ dB gain induced by channel narrowing, 
the guard-band $g$ (Hz) and the FD capability assumed before, the total capacity
$C_{fd:2}$ achieved within $B$ is given by Eq.~\ref{eq:1:n-fd}.
\begin{eqnarray}
 C_{fd2} &=& 4\Big(\frac{B-g}{2}\Big)\log_2\big(1 + 10^{(SNR_{dB}+3)/10}\big)\label{eq:1:n-fd}
\end{eqnarray}

\section{FD Wi-Fi MAC Protocol Capacity Upper-Bound} \label{sec:model}
In this section we characterize the ideal condition to derive the capacity upper-bound
of a Wi-Fi compliant FD MAC protocol. Then, we present a model to assess such capacity
under both the $1$:$1$ and $1$:$N$ MAC design guidelines.

\subsection{Ideal Condition for Wi-Fi Compliant FD MAC protocols} \label{subsec:idealconditions}
To keep Wi-Fi compliance, a MAC protocol shall follow the CSMA/CA access method.
In the context of FD radios, this means that \emph{at least} the primary transmission
initiates following a typical exponential back-off procedure. Since CSMA/CA is
half-duplex by nature, some additional mechanism is required to admit a collision-free
secondary transmission. The resulting time overhead to coordinate such a secondary transmission
(i.e. $\Delta_t$) is the key reason why MAC protocols' performance falls well below the FD 
gains~\cite{xie-doesfddouble-infocom-2014}. Therefore, under an `ideal FD condition', an
Wi-Fi compliant MAC protocol maximizes the FD gain utilization by minimizing the time 
overhead $\Delta_t$.

A naive way of characterizing the `ideal FD condition' is assuming $\Delta_t$$=$$0$ i.e.
$t_{st}$$=$$t_{pr}$. This implies that the same backoff number is shared without overheads
by a pair of arbitrary nodes at the beginning of each time slot. This is a too strong assumption
for our scenario because conflicts with the random uniform behavior of the CSMA/CA
backoff procedure. A reasonable alternative for this consists in assuming that \emph{the PR always
has a data frame enqueued to the PT}. In our scenario this means that the minimum $\Delta_t$
corresponds to the time interval the AP needs to start the secondary transmission just after
processing the incoming primary transmission's header $H_1$. A prior work has shown an
AP can manage to do that in real-time~\cite{jain-fd-mobicom11}.
The whole process is illustrated in Fig.~\ref{fig:fdchannel}. In the figure, an arbitrary
STA starts a primary transmission to the AP at the time instant $t_0$ upon winning a CSMA/CA 
contention (not illustrated). After receiving and processing $H_1$, the AP fetches a data frame
and starts a secondary transmission to the corresponding STA at the time $t_2$. This defines
the minimum $\Delta_t$, which corresponds to $t_2$$-$$t_0$$>$$0$ in the figure. Note, however, 
that FD becomes profitable only at $t_3$, the time at which useful data starts being transferred. 
To avoid collisions due to hidden terminals, both transmissions have to be finished 
simultaneously~\cite{singh-contraflow-wiopt-2011}, then the maximum secondary payLoad $L_2$
(bytes) for the capacity upper-bound is dimensioned accordingly. The other parameters on the
Fig.~\ref{fig:fdchannel} are helpful for the capacity model, as we explain next.

\subsection{Capacity Limit Model}\label{subsec:capacitymodel}
To compute the capacity limit of CSMA/CA under the ideal FD condition for each design guideline,
we refer to the IEEE 802.11 capacity model proposed by Bianchi~\cite{bianchi-jsac-2006}.
The model is twofold. Firstly it computes the probabilities $\tau$ 
and $p$ that a CSMA/CA station transmits and collides, respectively. These
probabilities are computed in the same way for our scenario, since the
STAs contends for primary transmissions just as in half-duplex CSMA/CA.
The second part of the model consists in a expression that computes the throughput
for IEEE 802.11 WLANs regardless of the channel access mode. More precisely,
the model computes the saturation capacity $S$ given both the payload carried per 
transmission and the time duration of each possible event in the channel. 

To assess $S$ assuming an  FD channel, we need firstly to characterize the possible events
related to a primary transmission at the beginning of a time slot. In our case they correspond
to same possible events of a CSMA/CA half-duplex channel, namely, `success', 
`collision' or `absent' (empty slot). These events happen with probabilities $P_{s}$,
$P_{c}$ and $P_{i}$ and take $T_s$, $T_{c}$ and $T_{i}$ absolute time units (e.g. $\mu$s),
respectively. Of these,  $T_{i}$ is obtained straightforwardly from the standard 
waiting slot time \cite{ieee80211-12}. Moreover, only the first event carries an expected amount 
of useful payLoad, that we denote as $E[L]$.

\subsubsection{Probabilities of channel events}
To compute $P_{s}$, $P_{c}$ and $P_{i}$, recall that each one of all $n$ STAs
does transmit with probability $\tau$ and \emph{does not} with probability $(1$$-$$\tau)$.
Thus, channel is idle with probability $P_{i}$$=$$(1$$-$$\tau)^n$. 
A primary transmission succeeds if only a single STA transmits and the remainder $(n$$-$$1)$ 
STAs remain silent, what happens with probability $\tau(1$$-$$\tau)^{n-1}$. Since each of the
$n$ STAs has the same chance to succeed $P_s$$=$$n\tau(1$$-$$\tau)^{n-1}$. A collision
happens if the channel is not idle and, at the same time, a primary transmission does not
succeed i.e. $P_{c}$$=$$(1$$-$$P_i)(1$$-$$P_{s})$. 

\subsubsection{Duration and payload of a successful primary transmission}
Let $H_1$ and $L_1$ be the PHY-MAC headers and payLoad sizes of a primary transmission,
respectively. Similarly, $H_2$ and $L_2$ have equivalent meaning for a secondary transmission,
as illustrated on Fig.~\ref{fig:fdchannel}. Also, let $T_H$ and $T_L$ be the time taken to 
transmit $H_1$ (or $H_2$) and $L_1$ under given control and data rates, respectively. 
 Denoting as $SIFS$ plus $T_{ACK}$ the total IEEE 802.11 standard time
needed to acknowledge a data frame and $\delta$ as the propagation time of each frame,
the overall duration of a successful primary transmission is given by Eq.~\ref{eqn:ts}. 
Note that $T_s$ also comprises $DIFS$ i.e. the minimum Wi-Fi standard time interval all STAs
must wait before assuming channel is idle again and restarting the CSMA/CA count-down.
\begin{eqnarray}
  T_s = T_H + T_{L}+ \delta + SIFS + T_{ACK} + \delta + DIFS \label{eqn:ts}
\end{eqnarray}
\input{figures/ideal-fdcsma}
As one can also see on Fig.~\ref{fig:fdchannel}, under the ideal FD condition, a dual-link 
comprises two data frame transmissions. Therefore, the total expected payload carried within
the channel event `success' is defined as $E[L]$$=$$L_1$$+$$L_2$. Note that
$L_2$$=$$L_1$$-$$f_L(H_2)$, where $f_L(H_2)$ is the amount of useful payload that the secondary
transmission's data rate could send during the time interval comprising fetching and transmitting
$H_2$ (i.e. $[t_1,t_3]$ on Fig.~\ref{fig:fdchannel}).

\subsubsection{Duration of a collision} To detect a collision, the PT starts its timer
just after pushing the last symbol header into the channel. As soon as PT detects
an incoming symbol, it stops the timer and finishes receiving the whole incoming signal. 
If the received signal does not correspond to $H_2$ as expected or, alternatively, no signal 
is detected before the timer expires, then PT stops transmitting. The maximum timer estimation
comprises $T_H$, the header propagation time and the overhead on PR to start the secondary
transmission appropriately. In~\cite{jain-fd-mobicom11} authors report an overhead of $11\mu$s to 
start an FD Wi-Fi like transmission in real-time.

\subsubsection{MAC guidelines saturation throughput} 
The FD CSMA/CA capacity formula $S$ (Eq.~\ref{eqn:s}), comes from the ratio between the payload
and  the time duration associated to each possible event in the channel. 
\begin{eqnarray} \label{eqn:s}
  S &=& \frac{P_s(L_1+L_2)N}{P_sT_s + P_{c}T_{c} + P_{i}T_{i}}
\end{eqnarray}
Each value in the ratio are weighted by the corresponding channel event probability. This formula
stands for both design guidelines.
The difference is that $N$$=$$1$ for the $1$:$1$ design. Hence, under the ideal  FD condition, each
CSMA/CA round triggers two transmissions across the whole channel. With the $1$:$N$ design, $N$$>$$1$ 
and each CSMA/CA round triggers $2$$\times$$N$ narrow-channel transmissions under the same ideal 
condition. Also, all timing parameters are rescaled according to channel width just as the IEEE 802.11
standard mandates~\cite{ieee80211-12}.

\section{Results}\label{sec:results}
In this section we report results achieved by both $1$:$1$ and $1$:$N$ guidelines at 
and below the MAC layer. We also report the half-duplex performance for comparison purposes.
 
\subsection{Novel Capacity Limit Below the MAC Layer}
In Fig.~\ref{fig:1:Nshannoncapacity} we plot the capacity upper-bound for 
the $1$:$1$ and the $1$:$N$$=$$2$ guidelines across different SNRs
(Eqs.~\ref{eq:shannondB} and~\ref{eq:1:n-fd}, respectively). The total
bandwidth is $B$$=$$20$ MHz so $1$:$N$ corresponds to two $10$ MHz channels. 
Each $10$ MHz channel is separated by a
guard-band $g$$=$$100$ KHz, what can be achieved by actual filters 
e.g.~\cite{chintalapudi-wifinc-usenix-2012}. We also plot the half-duplex 
capacity for comparison purposes (Eq.~\ref{eq:shannon}). As widely known, the gain
of any FD radio is bounded by $2\times$ the half-duplex capacity. However, the SNR
gains induced by channel narrowing breaks this currently prevalent gain
even paying a $100$ KHz guard-band overhead. We verified the statement still holds
for a $g$ up to about $1.1$ MHz.
\input{figures/shannon}

To investigate whether such theoretical results preserves 
in practice,  we propose a proof-of-concept study based on a pair of Ettus USRP B210
Software Defined Radio (SDR) platforms. Each radio is equipped with one antenna for 
transmission and one for reception. We compare a single $10$ MHz channel against 
two $5$ MHz channels. An \emph{ideal} FD radio doubles capacity by entirely releasing 
the bandwidth for reception while transmitting. To mimic such behavior, we rely on an
\emph{out-of-band} FD test. Thus, in both scenarios, each radio has $10$ MHz channel 
dedicated for reception and another $10$ MHz for transmission, being these channels $60$ MHz 
away from each other. For each case, we set the highest modulation the IEEE 802.11 
standard mandates under a Received Signal Strength Indication (RSSI) of $-80$ dBm i.e. QPSK $3/4$ for 
$10$ MHz and $16$-QAM $1/2$ for $5$ MHz~\cite[Table 18--14]{ieee80211-12}. This yields data rates 
of $9$ and $6$ Mbps, respectively. We produce Wi-Fi signals based on the \texttt{gr-ieee80211}
GNURadio module~\cite{grieee80211-bastian-13} and measured all bytes transferred
through saturated links. Since SDR experiments are dramatically affected by CPU load
and FD doubles such processing demands, we assess the half-duplex link from the best 
FD link. For each experiment we gather as much sample as needed to calculate
mean throughput with a confidence of $95\%$ and a relative error $<5\%$, 
following the statistical procedures of \cite{ewing-akaroa2-1999}.
\input{figures/usrp-shannon}

From the plots on Fig.~\ref{fig:1:usrpshannon} one can see our theoretical statement holds 
in practice. Our design improves over half-duplex about $2.2\times$.
For all cases the throughput is dramatically lower in comparison to theory 
mostly because of the large latency introduced by the (half-duplex) USB connection 
between USRP and PC. Finally, we recognize that the capacity of actual \emph{in-band}
FD radios is strictly less than $2\times$ half-duplex's because of residual self-interference.
\emph{However, our findings suggest that the gains claimed by single-band 
FD radio proposals might be underestimated}. For instance, we believe the best currently 
reported result -- $1.87\times$ in an $80$ MHz channel with an RSSI of 
$P_R$ dBm~\cite{bharadia-fullduplex-sigcomm-2013} -- could be even better if performed over 
two $40$ MHz FD channels (with appropriate filters/guardbands) set to the densest Wi-Fi
modulation scheme supported under $P_R$ dBm.

\subsection{Novel Capacity Limit at the MAC Layer}
To check whether the PHY layer improvement \emph{scales} at the MAC layer we assess
the capacity upper-bound of the FD CSMA/CA under both $1$:$1$ and $1$:$N$ designs. 
The numerical results are computed in accordance with the section
\ref{subsec:capacitymodel}.
We also report the half-duplex results under both the standard access modes namely,
the $2$-way (i.e. DATA followed by ACK) and the $4$-way
handshakes (Request-to-Send/Clear-to-Send, RTS/CTS-DATA-ACK). We assume a
propagation time of $\delta$$=$$1\mu$s, a bandwidth of $B$$=$$20$ MHz and $N$$=$$2$
(i.e. two $10$ MHz channels for $1$:$N$).
All other timing parameters are set according to the IEEE 802.11a best-effort traffic
class.

We verify that the FD MAC protocols outperform the half-duplex CSMA/CA across 
different data rates and frame payload sizes. Due to space constraints, on
Fig.~\ref{fig:mac-comparison48Mbps} we only report results for data 
rate of $48$ Mbps in $20$ MHz channels. This implies in at least $27$ Mbps for $10$ MHz
channels~\cite{ieee80211-12}. Similarly, for these respective channel widths, we set 
control rates to $18$ Mbps and $12$ Mbps and MAC payload to $788$ bytes. Larger payloads
dramatically damages $2$-way half-duplex performance upon collisions, specially as network
grows
(Fig.~\ref{fig:mac-comparison48Mbps}). 
The $4$-way handshake mitigates that by preceding data transmission with smaller RTS
frames but the overall handshake slows all successful transmissions.
In turn, with FD only a very small part of the primary transmission's payload
is exposed to collision. This happens with no penalty to successful transmissions. 
In addition to these abilities, the poor half-duplex performance over an increasing
number of nodes causes the $1$:$1$ FD CSMA/CA to be as higher as $2$$\times$ the 
half-duplex performance (as of $\approx$$290$ nodes on Fig.~\ref{fig:mac-comparison48Mbps}).
However, as one can also see on Fig.~\ref{fig:mac-comparison48Mbps}, such gains can be 
improved by conforming the FD CSMA/CA to the $1$:$N$ design.  Indeed, the MAC gain under
the $1$:$N$ design closely approaches the PHY layer improvement we report in this section, 
and keeps scaling over nodes.
Moreover, the channel orthogonality exploited by the $1$:$N$ design enables higher spatial 
reuse.  Hence, the number of FD opportunities in the best-case increase from $2$ to $2$$\times$$N$.
Although non-exhaustive, these results represent an unprecedented step towards the 
scalability of FD gains in single-cell WLANs.
\input{figures/mac-66dBm}

\section{Conclusion and Future Work}\label{sec:conclusion}
In this work we study the capacity limits of single-cell FD WLANs. 
We inquire what prevents current Wi-Fi compliant FD MAC protocols 
to fully profit from the theoretical double of throughput leveraged by FD radios.
In addition to the overheads \emph{at} the MAC layer, we realize this is also  
explained by the capacity bound imposed \emph{below} the MAC layer. Thus, we 
propose a design categorization based on which MAC protocols are classified
according to the way they `see' the FD bandwidth. In this sense, we identify
current FD Wi-Fi MAC protocols are classified into what we refer to as the $1$:$1$ 
design guideline, meaning they `see' the FD bandwidth through a \emph{single} PHY layer.
With this, MAC performance is bounded by a pair of transmissions in the channel. Instead,
under the $1$:$N$ design guideline we advocate, MAC protocols `see' the FD bandwidth
through $N$$>$$1$ orthogonal narrow-channel PHY layers. Based on theoretical 
results and software defined radio experiments, we show it is possible to outperform
the current assumed FD capacity gain at and below the MAC layer. 
To benefit from this novel more-than-doubling improvement, in future works we plan to 
design novel mechanisms that exploit the spatial reuse opportunities enabled by the $1$:$N$
design guideline. Also, we intend to study the $1$:$N$ design together the MIMO technology. 
%
\begin{spacing}{0.9}
\bibliographystyle{ieeetr} 
\bibliographystyle{abbrv}
\bibliography{refs}  %
\end{spacing}

\end{document}

%% file: figures/spatial-reuse.tex
\begin{figure}[t]
\includegraphics[scale=0.35]{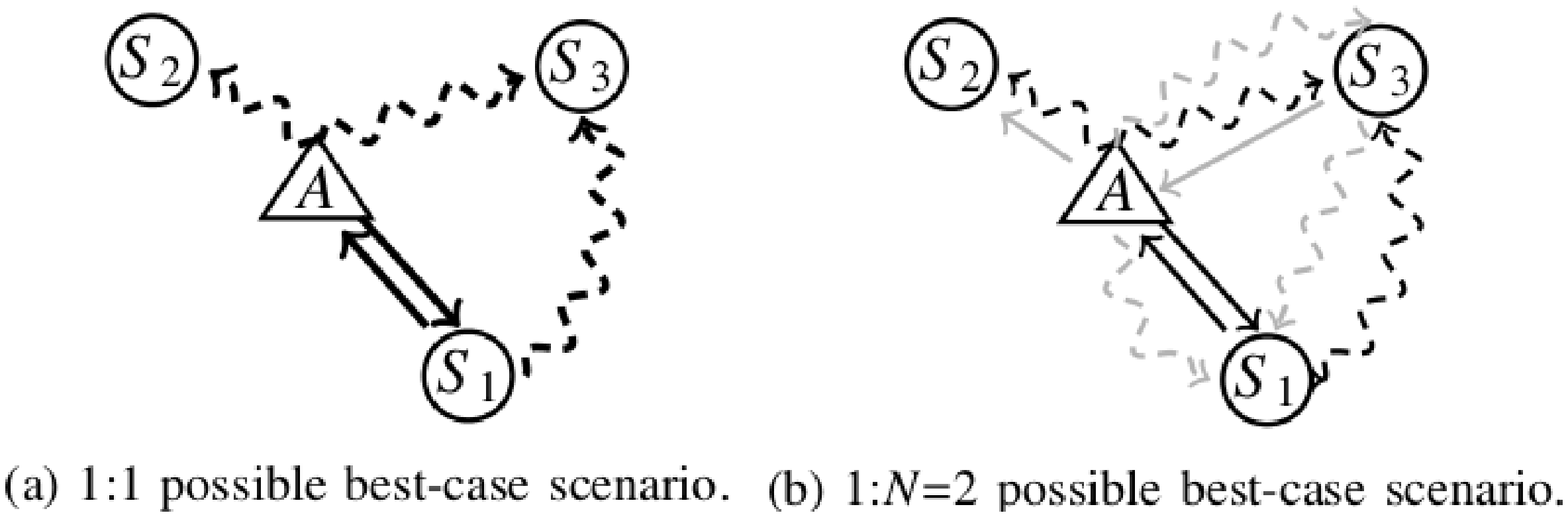}
 \caption{Best-case comparison: Under $1$:$N$ (b), the number
          of dual-links (couple of solid straight arrows) outperforms $1$:$1$ (a) by 
          a factor of $N$. Channel orthogonality (gray and black 
          colors) overcomes interference (dashed waved arrows) to increase spatial reuse.}
 \label{fig:spatial-reuse-comparison}
\end{figure}

%% file: figures/psd.tex
%
%
%

\begin{figure}[t]
\centering
\includegraphics[scale=0.45]{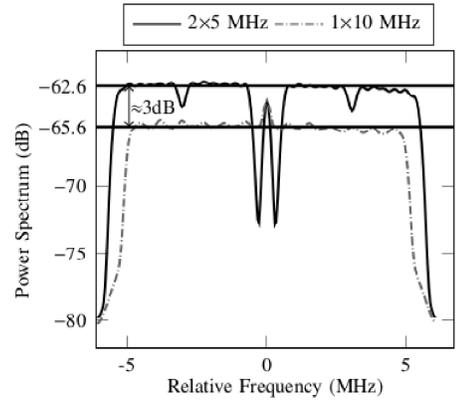}
   \caption{Each concurrent $5$ MHz-wide channel outperforms a single $10$ MHz channel about $3$ dB even
            under the same output power.}
   \label{fig:psd}
\end{figure}

%% file: figures/ideal-fdcsma.tex
\begin{figure}
\includegraphics[scale=0.5]{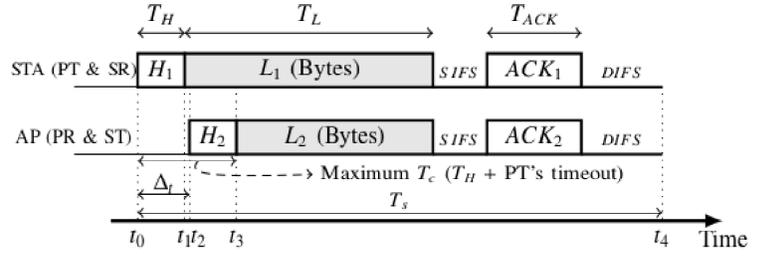}
\caption{Ideal FD condition for the performance of an Wi-Fi compliant FD MAC protocol.
         The AP (PR) always has a frame enqueued to the STA (PT). 
         At time $t_2$ the AP (ST) starts sending a data frame to the STA (SR) upon
         receiving and processing the primary transmission header (during [$t_0$,$t_2$]).}
\label{fig:fdchannel}
\end{figure}

%% file: figures/shannon.tex
\begin{figure}[t]
\centering
\includegraphics[scale=0.5]{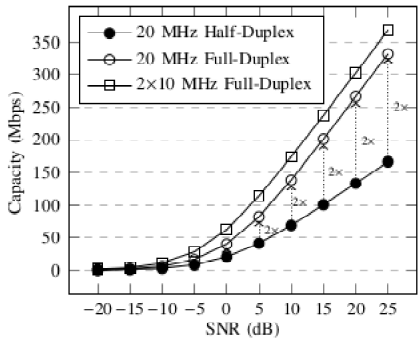}
   \caption{Maximum capacity delivered to the MAC layer by two $10$ MHz 
            FD Wi-Fi channels ($1$:$N$ design, $N$$=$$2$) against full and half duplex 20 MHz channels. 
            The $1$:$N$ design  more than doubles half-duplex capacity paying a guard-band
            overhead $g$$\leq$$1.1$ MHz (plots for $g$$=$$100$ KHz assuming actual filters 
            e.g.~\cite{chintalapudi-wifinc-usenix-2012}).}
   \label{fig:1:Nshannoncapacity}
\end{figure}

%% file: figures/usrp-shannon.tex
\begin{figure}[t]
   \centering
\includegraphics[scale=0.5]{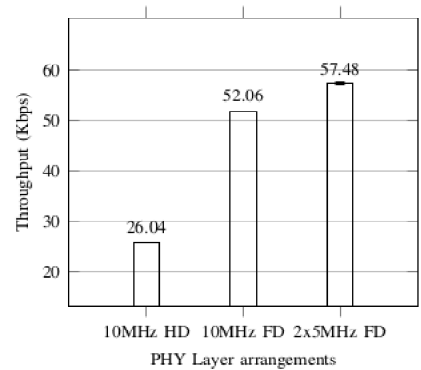}
   \caption{Capacity of ideal Full-Duplex (FD) radios mimicked by (out-of-band)
            FD experiments on USRP platforms. Two $5$ MHz FD channels
            outperform a single FD $10$ MHz channel i.e. more than doubles Half-Duplex's (HD) capacity.}
   \label{fig:1:usrpshannon}
\end{figure}

%% file: figures/mac-66dBm.tex
\begin{figure}[t!]
\centering
\includegraphics[scale=0.5]{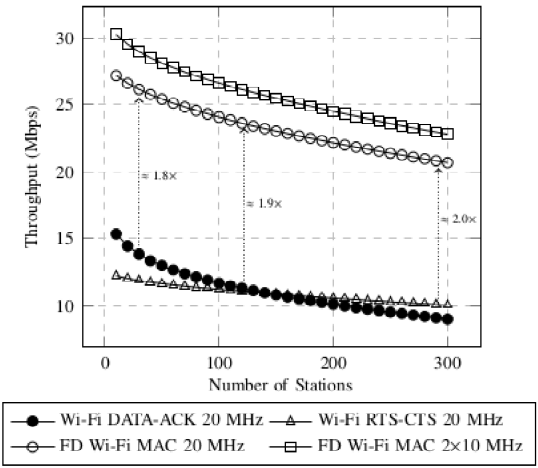}
 \caption{Capacity of the $1$:$1$ (single-band) FD Wi-Fi MAC protocol improves under the $1$:$N$ design guideline ($N$$=$$2$).}
 \label{fig:mac-comparison48Mbps}
\end{figure}